# Optimization Studies for the Long-Baseline Neutrino Facility at Fermilab         Fermilab-Conf-22-710-AD

Igor Rakhno,* Nikolai Mokhov,# Igor Tropin,§ Sergei Striganov ∆ †

*Fermi National Accelerator Laboratory, P.O. Box 500, Batavia, Illinois 60510-5011, rakhno@fnal.gov
#Fermi National Accelerator Laboratory, P.O. Box 500, Batavia, Illinois 60510-5011, mokhov@fnal.gov
§Fermi National Accelerator Laboratory, P.O. Box 500, Batavia, Illinois 60510-5011, tropin@fnal.gov
†Fermi National Accelerator Laboratory, P.O. Box 500, Batavia, Illinois 60510-5011
∆ Deceased

## INTRODUCTION

The Deep Underground Neutrino Experiment and Long-Baseline Neutrino Facility (DUNE-LBNF) are under development at Fermilab since early 2010s [1]. At present, the work is being performed towards a comprehensive review conducted by US Department of Energy (DOE)—the Critical Decision 2 (CD-2)—that is planned to take place in the middle of 2022. The primary scientific objectives of DUNE are to carry out a comprehensive investigation of neutrino oscillations to test CP violation in the lepton sector, determine the ordering of the neutrino masses, and to test the three-neutrino paradigm (electron, muon and tau neutrino). The LBNF will provide a 120-GeV proton beam on a neutrino production target utilizing a new 800-MeV superconducting Linac which is expected to be completed in 2027 [2]. The neutrino beamline, which utilizes a target and horn systems, decay pipe, hadron absorber and other systems, is a core component of the LBNF. At present—as a result of numerous iterations—there exists an optimized design with a 1.5-m graphite target and focusing system consisting of three horns. Further optimization energy deposition and radiological calculations are performed towards CD-2 and beyond. This paper describes results of the most recent MARS15 [3] optimization studies.

## MARS15 Model of the LBNF Beamline

A detailed model of the entire LBNF complex has been developed with the MARS15 code [3] using a built-in three-dimensional MAD-X based beam-line builder [4] and ROOT geometry [5] that provides great flexibility when building complex geometry structures. The model features all the major components of the facility shown in Fig. 1: (i) primary beamline that starts at the beam extraction system in the Main Injector and delivers the proton beam to the target hall; (ii) the target hall itself with the target station and corresponding engineering infrastructure; (iii) decay channel; (iv) hadron absorber with a cooling system, muon detectors and other equipment in the service building; and (v) muon kern downstream of the hadron absorber service building. The model shown in Fig. 1 employs the following major color coding: light blue, green, turquoise, grey and yellow correspond to air, glacial till, dolomite, concrete and helium, respectively. Fragments of the entire model that show the target hall and hadron absorber complex in more detail are shown in Figs. 2 and 3, respectively. Magnetic field distributions in the magnets and focusing horns were accounted for as well. Thorough testing of the model has been performed to make sure it correctly predicts neutrino beam delivery to both the Near Detector at Fermilab and Far Detector in South Dakota. Also, comparisons with neutrino flux predictions based on Geant4 [6] simulations, performed for the same model, revealed a good agreement between MARS15 and Geant4. In this case, the model transfer from MARS15 to Geant4 has been performed using the GDML format.

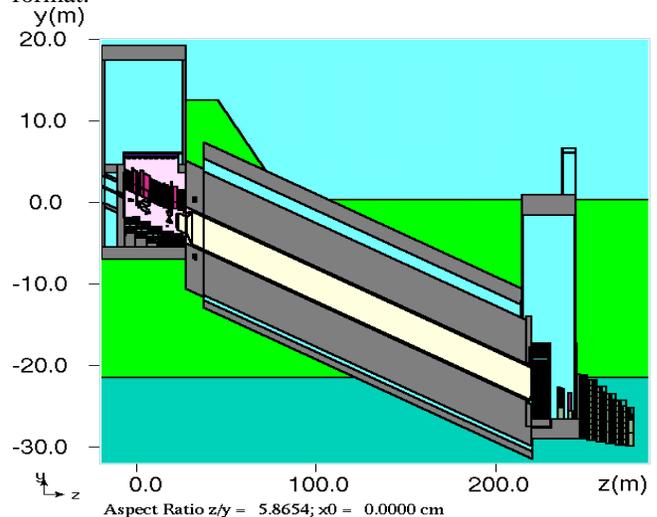

Fig. 1. Elevation view of the major components of LBNF at Fermilab: target hall, decay channel and hadron absorber complex with muon kern.

## RESULTS

All the Monte Carlo calculations performed so far for the LBNF can be combined into several groups: (i) energy deposition calculations that help predict heating for various beamline components; (ii) prompt dose rate distributions near the beamline and inside the service buildings atop that are needed by themselves and to predict lifetime for components with organic materials like epoxy, G10 insulation *etc*; (iii) surface and ground water activation around the facility; (iv) air activation; (v) activation of the beamline components; (vi) residual dose distributions around the beamline that are important from the standpoint of beam-off maintenance; (vii) prompt dose distributions outside the facility (above the berm); (viii) calculations for the muon kern; (ix) calculations of neutrino fluxes and spectra in both the Near and Far Detectors. Such a separation into several groups means that corresponding calculations are performed—in most cases—separately. It is justified not only by the fact that the entire model is very large spatially and corresponding calculations are very time consuming. From physical standpoint the difference between the groups is due to the fact that the different energy thresholds should be applied to particles that play major role in every single case.

Distribution of the beam-induced power dissipation in the major LBNF systems for the planned 2.4 MW on target operation is shown in Table I. Several beam accident scenarios were also studied. The major results for each of the major LBNF system are described below.

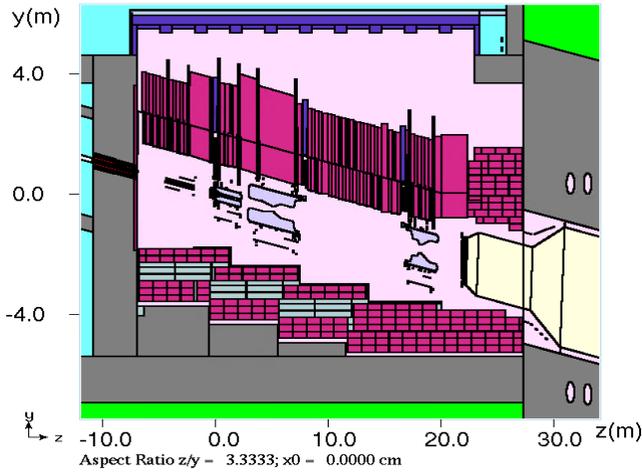

Fig. 2. A fragment of the model that shows the LBNF target hall (elevation view). The light violet color corresponds to nitrogen used instead of the air in order to mitigate corrosion issues. The upstream end of the target itself is located at y = 0 and z = 0.

*Primary Beamline*

Accidental beam loss represents a major threat to the primary beamline. It has been shown in several detailed Monte Carlo simulations (followed by thermal analyses) that even a single lost beam pulse with grazing angle above 3 mrad can lead to the local beam pipe meltdown. At grazing angles not exceeding 1.5 mrad, the beam pipe can survive after two consecutive beam pulses lost at the same location. An accident with the normal beam incidence (for instance, at an aperture transition) can destroy the corresponding magnet. At the Fermilab site boundary, prompt dose caused by an accidental beam loss at the beamline apex contributes less than $10^{-7}$ mSv/pulse. A calculated distribution of prompt dose around LBNF due to an accident at beamline apex is shown in Fig. 4. The fluence-to-dose conversion factors from [7-10] are used.

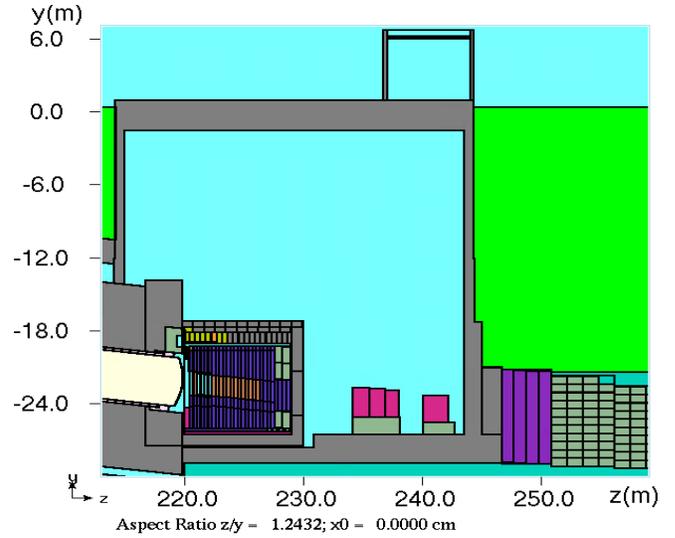

Fig. 3. A fragment of the model that shows the hadron absorber complex in more detail (elevation view). The absorber core consists of aluminum and steel blocks surrounded with concrete shielding.

TABLE I. Power dissipation in the LBNF systems

| System | Power (kW) |
|---|---|
| Target station | 1132 |
| Decay channel | 555 |
| Hadron absorber complex | 522 |
| Generated neutrino power | 45 |
| Others | 146 |
| Total | 2400 |

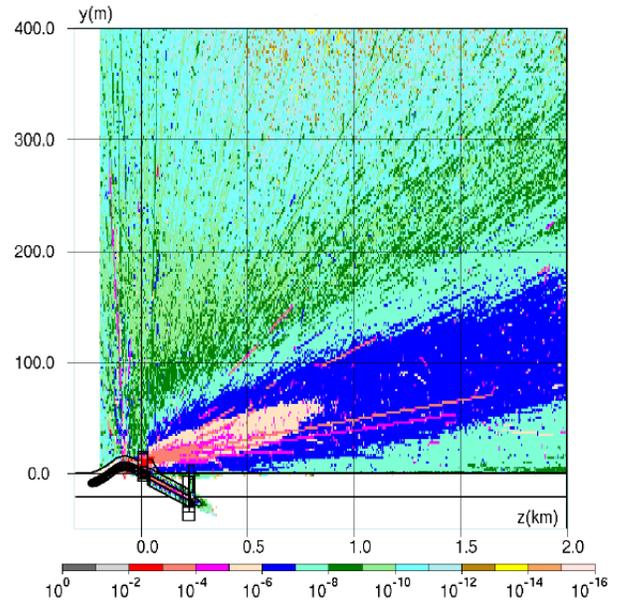

Fig. 4. A calculated prompt dose (mSv/pulse) distribution around LBNF for an apex accident scenario. Neutron energy threshold of 100 keV has been used in these calculations instead of a default thermal neutron energy used in all other cases.

*Target Station*

Numerical studies of the target length, lateral shape and dimension, support structure, gaps, baffle and bafflette parameters have been performed. The major goal of these studies was reducing peak energy deposition in the hottest regions of the hadron absorber. Current design focuses on the RAL's 1.5-m graphite target design. In the target hall, the beam power is deposited mostly in the cooling panels, steel shielding blocks as well as horns and stripline (480, 355 and 169 kW, respectively) [11].

*Decay Pipe*

The following quantities have been calculated for the double-walled decay pipe: (i) heat load in its components; (ii) tritium production rate in the beam pipe and concrete shielding around it; (iii) energy deposition in the downstream window and frame and in the concrete shielding; (iv) surface water activation around the beam pipe shielding. The beam power is deposited mostly in the double-walled steel decay pipe and concrete shielding (341 and 185 kW, respectively) [11]. It has been shown that temperature in the window downstream of the decay pipe is within tolerable limits. Spatial distributions of the tritium production rate in the pipe were used in a subsequent analysis of the tritium diffusion. The predicted surface water activation has been shown to comply with regulatory requirements.

*Hadron Absorber*

In the current design, the hadron absorber core is a uniform structure. The major goal of multiple performed iterations was keeping maximum temperature in its aluminum core below the tolerable level of 100º C. The goal has been achieved by means of using various beam spoiling techniques such as optimizing baffle and bafflette lengths. It allowed us to significantly reduce the fraction of (almost) un-scattered incoming high-energy proton beam. Also, optimization of water-cooling channels inside the hadron absorber has been performed.

In the hadron absorber, the beam power is dissipated mostly in the aluminum spoiler and core as well as absorber steel (229 and 244 kW, respectively) [11].

Another essential topic related to the hadron absorber complex is radiological studies. Various radiological quantities have been calculated: prompt and residual dose distributions around the absorber and the service building that accommodates it, surface and ground water activation around the hadron absorber complex, tritium production in steel and concrete, and air activation. A sophisticated splitting technique has been used for this deep-penetration shielding problem. It has been shown that the predicted prompt and residual dose rates in the upper part of the service building are below the regulatory limits that correspond to areas of unlimited occupancy.

A calculated distribution of prompt dose in the LBNF hadron absorber complex is shown in Fig. 5. One can see that the predicted prompt dose in the upper part of the service building—directly above the absorber—is very low and does not exceed $5\times10^{-7}$ mSv/hr. The prompt dose in the absorber hall at the level of the absorber top does not exceed 350 mSv/hr, and immediately downstream of the absorber the maximum prompt dose is about $8\times10^4$ mSv/hr. The peak prompt dose rate on top of the absorber, predicted on the downstream end of the concrete shielding around the absorber steel shielding, is about 40 mSv/hr. Also, in the upper service building above the absorber the maximum predicted residual dose on the floor is about $10^{-11}$ mSv/hr for a 100-day irradiation and 4-hr cooling scenario.

For air activation estimates, hadron flux above 30 MeV has been calculated. In the hottest spot of the absorber hall air—between the decay pipe downstream window and face of the absorber—the average flux is about $3.2\times10^9$ cm$^{-2}$ s$^{-1}$, which corresponds to a very low air activation level that is in compliance with current regulatory requirements.

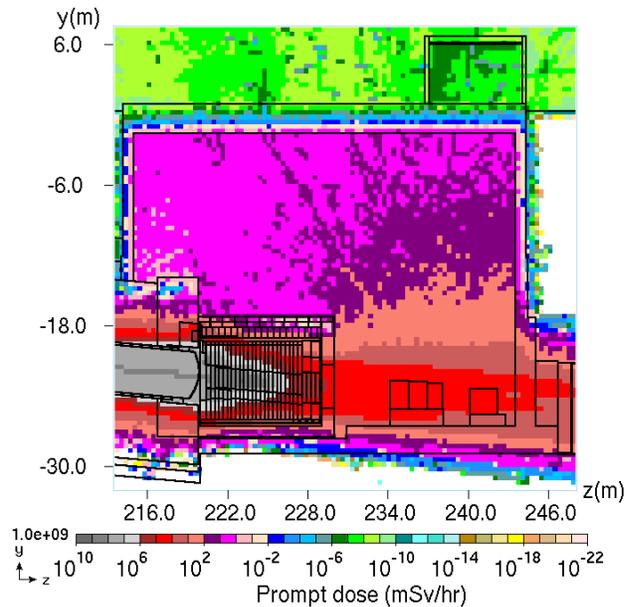

Fig. 5. A calculated distribution (elevation view) of prompt dose around the hadron absorber complex for normal operation at 2.4 MW.

For ground water activation, hadron flux above 30 MeV has been calculated outside the absorber hall shielding and muon kern. The flux has been averaged—over a layer as thick as 1 cm—immediately outside the concrete and steel. The 1-cm thick layer is a combination of layers adjacent to various sections of the concrete floor and walls and steel blocks. Also, these layers contain both rock (dolomite) and glacial till. The total volume of this 1-cm layer is $3.8\times10^7$ cm$^3$. The calculated averaged hadron flux above 30 MeV over the entire layer is about 14 cm$^{-2}$s$^{-1}$. Volume of the 1-cm layer around the muon kern itself and the averaged hadron flux in this layer are 6.1×10$^6$ cm$^3$ and 33 cm$^{-2}$s$^{-1}$, respectively. As long as ground water activation is concerned, two radionuclides have highest production rates and longest half-lives, namely $^3$H and $^{22}$Na. Further analysis revealed that the predicted amounts of $^3$H and $^{22}$Na produced in ground water around the Hadron Absorber Complex are below detectable levels.

*Neutrino Fluxes in DUNE Far Detector*

Precision measurements of neutrino fluxes is one of the primary scientific goals of DUNE. In order to ensure an appropriate beamline design and optimization, the computer codes used for that purpose must undergo a verification in terms of their predictions. To that end, two independent studies of predicted neutrino fluxes at DUNE far detector have been performed using G4LBNF [12] and MARS15 [3] codes. The calculated muon neutrino fluxes at the far detector are shown in Fig. 6. One can see that in the energy range of major interest—up to neutrino energy of a few GeV—both the codes predict quite similar results.

Regarding neutrinos that enter the hadron absorber and detectors, their calculated power breakdown is presented in Table II.

Table II. Power of neutrinos that enter the hadron absorber and detectors

| System | Power |
|---|---|
| Hadron Absorber Complex | 11.7 kW |
| Near Detector | 2.8 kW |
| Far Detector | 3.2 mW |

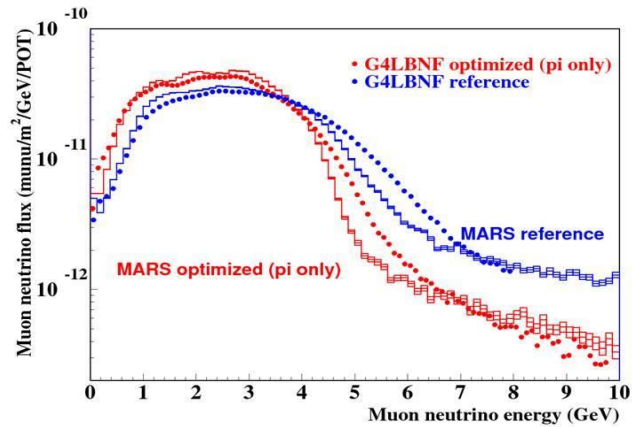

Fig. 6. Muon neutrino fluxes calculated at DUNE Far Detector for reference and optimized design using G4LBNF [12] and MARS15 [3] codes. For the optimized design, at high energies neutrino generation modeling has been done for pion decays only. Neutrino generation due to kaon decays at high energies has been turned off because this study has been focused on neutrino energies below 6 GeV.

**ENDNOTES**

As a result of comprehensive studies including the energy deposition and radiological calculations, the neutrino beamline design has been advanced to a significant maturity level. This includes all the major subsystems such as primary beamline, target station, decay channel and hadron absorber.

This work is supported by Fermi Research Alliance, LLC under contract No. DE-AC02-07CH11359 with the U.S. Department of Energy.
This research used Fermigrid of the Fermi National Accelerator Laboratory and an ALCC allocation at the Argonne Leadership Computing Facility, which is a DOE Office of Science User Facility supported under Contract DE-AC02-06CH11357.